\pdfoutput=1
\documentclass[11pt]{article}
\usepackage[switch]{lineno}
\usepackage[review]{acl}

\usepackage{times}
\usepackage{latexsym}
\usepackage{natbib}
\usepackage{enumitem}
\usepackage[T1]{fontenc}
\usepackage[utf8]{inputenc}

\usepackage{microtype}

\usepackage{inconsolata}

\usepackage{graphicx}
\usepackage{multirow} % Add this to your preamble

\title{Probing Audio-Generation Capabilities of Text-Based Language Models}

\author{
    Arjun Prasaath Anbazhagan\thanks{These authors contributed equally to this work.} \hspace{0.7em} 
    Parteek Kumar\footnotemark[1] \hspace{0.7em}
    Ujjwal Kaur\footnotemark[1] \hspace{0.7em} \\ 
    \bfseries Aslihan Akalin\thanks{Research mentor.} \hspace{0.7em} 
    Kevin Zhu\footnotemark[2] \hspace{0.7em} 
    Sean O'Brien\footnotemark[2] \\
    \texttt{Algoverse AI Research} \\
    \texttt{
        \href{mailto:arjunanbazhagan2026@u.northwestern.edu}{arjunanbazhagan2026@u.northwestern.edu} \hspace{1em}
        \href{mailto:\{kevin, asli, sean\}@algoverseairesearch.org}{\{kevin, asli\}@algoverseairesearch.org}
    }
}

\date{}

\begin{document}
% \linenumbers
\maketitle

{\makeatletter\acl@finalcopytrue}

\begin{abstract}
How does textual representation of audio relate to the Large Language Model's (LLMs) learning about the audio world? This research investigates the extent to which LLMs can be prompted to generate audio, despite their primary training in textual data. We employ a three-tier approach, progressively increasing the complexity of audio generation: 1) Musical Notes, 2) Environmental Sounds, and 3) Human Speech. To bridge the gap between text and audio, we leverage code as an intermediary, prompting LLMs to generate code that, when executed, produces the desired audio output. To evaluate the quality and accuracy of the generated audio, we employ FAD and CLAP scores. Our findings reveal that while LLMs can generate basic audio features, their performance deteriorates as the complexity of the audio increases. This suggests that while LLMs possess a latent understanding of the auditory world, their ability to translate this understanding into tangible audio output remains rudimentary. Further research into techniques that can enhance the quality and diversity of LLM-generated audio can lead to an improvement in the performance of text-based LLMs in generating audio.
\end{abstract}

\section{Introduction}
What do text-based large language models (LLMs) know about audio? Does the textual representation of sound entirely define an object or a thing? How accurately can textual representations capture auditory experiences? Can LLMs generate the sounds of musical instruments, like a flute, based solely on textual data? While text-based descriptions can outline physical characteristics -- such as a flute's structure or the rain and thunder of a thunderstorm -- are they enough to reproduce the sounds? We explore these questions by introducing a prompting-based approach toward audio generation using large language models (LLMs), investigating both human and machine understanding of auditory concepts. We also evaluate how large language models translate text into meaningful auditory representations.

Recent progress in LLMs has shown that models trained primarily on textual data can generate and comprehend auditory representations.

Various studies such as \citet{abdou2021languagemodelsencodeperceptual} and  \citet{patel2022mapping}, suggest that these models can process audio-related data and generate audio without explicit audio training. The development of methods such as Audio Spectrogram Transformers in \citet{gong21b_interspeech} demonstrates how LLMs can adapt to generate audio content, which could transform interactions with voice assistants. However, training such models involves significant computational costs, and the models often require vast amount of data. This highlights a key gap in the current field which this paper seeks to address:
\begin{enumerate}[itemsep=1pt, topsep=2pt]
    \item Can LLMs trained only on text be probed to generate audio?
    \item To what extent can the generated audio be comparable to its real-world counterpart?
    % to open source audio datasets.
\end{enumerate}

\noindent The paper’s main contributions are as follows.
\begin{itemize}[itemsep=1pt, topsep=4pt]
    \item \textbf{Audio Generation:} Demonstrating that LLMs trained solely on textual information can generate audio waveforms.
    \item \textbf{Audio Evaluation:} Assessing how well the generated audio matches real-world sounds.
\end{itemize}

\section{Literature Review}

\subsection{Audio and Transformer Models}
Transformers, originally designed for natural language processing, have been adapted for image analysis through the Vision Transformer and for audio signals \cite{verma2021audiotransformerstransformerarchitectureslarge}. This development has notably advanced deep learning in modeling long-term sequences, with key examples including the Music Transformer \cite{huang2018musictransformer} and AudioPaLM \cite{rubenstein2023audiopalmlargelanguagemodel}.

\subsection{Vision Check-up for Language Models}
\citet{sharma2024visioncheckuplanguagemodels}
evaluated language models trained on text-based inputs to check if they can understand images and generate them. They used code to convert between images and text. The findings revealed that while the model effectively learned basic objects and simple scenes, it had difficulty with complex scenes and detailed textures.

\subsection{Code Memorization}
%Our study also addresses the capability of LLMs in generating novel code for the audio synthesis task. Our findings revealed challenges in identifying readily available code that contains both spectrograms and corresponding audio similar to the one generated by the model. Investigation into the generated code revealed that LLMs do not merely retrieve pre-existing code snippets but construct audio signals from fundamental principles, incorporating various frequency modulations and acoustic parameters.

Significant research has explored whether LLMs simply reproduce code they have encountered in training. For instance, \citet{honarvar2023turbulence} demonstrates that LLMs can generate novel algorithms or approaches that differ from typical human-written code, suggesting that they do not simply reproduce training examples but synthesize new solutions. Similarly, \citet{zhang2024unseen} also suggest that when an unfamiliar or obfuscated prompt is passed to an LLM, it still can attempt to generate relevant code, albeit with reduced accuracy. These findings align with our observations. As the complexity of generation task increases, model performance declines, resulting in poor quality of generated audio. These results contribute to the ongoing discourse on LLM capabilities, suggesting that models possess a degree of creative problem-solving ability in code generation.

% \subsection{Program synthesis via LLMs}
% Pioneered by OpenAI’s Codex \cite{chen2021evaluatinglargelanguagemodels}, Github’s Copilot, Meta’s Code Llama \cite{rozière2024codellamaopenfoundation} etc., LLMs have been shown to possess exceptional coding capabilities. In this work, we explore the capabilities of large language models for audio generation through text-prompted code, carefully examining the diversity and realism of various text-only models such as GPT-4, GPT-4o, o1, and Llama 3.1 70B for programmable audio generation. 
% \subsection{Probing for Auditory Representations in Language Models}
%  \citeauthor{Ngo2024-rj} \citeyear{Ngo2024-rj} focused on how LLMs can retrieve correct text representations of objects from audio snippets using a linear probe trained on contrastive loss, but it does not explicitly state the robustness of model accuracy for different types of sounds (environmental sounds, musical sounds or speech) which is a focus of this research.

% Our paper is divided into two sections: Part I focuses on an LLM's capability to generate audio through 3 categories of sound.
% % ;
% % \begin{enumerate}
% %     \item Musical Notes
% %     \item Environmental Sound
% %     \item Human Speech
% % \end{enumerate}
% Part II focuses on evaluating these generated audio files using the Fréchet Audio Distance (FAD) and  Contrastive Learning Audio-Pretrained (CLAP) model.

\begin{figure*}[t]
    \centering
    \includegraphics [width=1\linewidth]{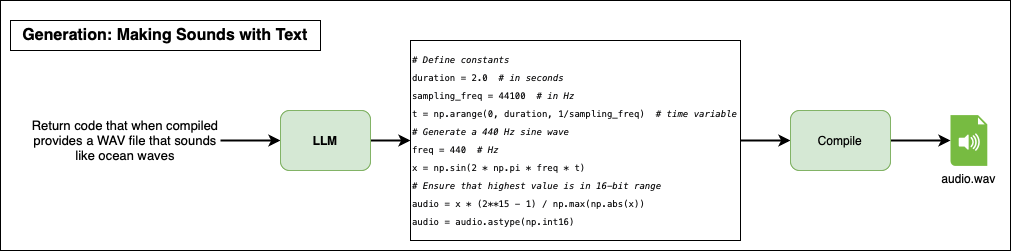}
    \caption{Pipeline for Language Model Audio Generation}
    \label{fig:part_1}
\end{figure*}

% \section{Datasets}

% To holistically evaluate an LLM's audio synthesis competence across audio categories, we used three different datasets.

% \subparagraph{Musical Notes:}
% For musical notes, we chose the NSynth Musical Notes \cite{nsynth} dataset which has 305,979 musical notes from 10 different instruments, 3 different sources with varying pitch and velocity and quality descriptions of the musical notes. From this dataset, we sampled 1500 audio files across 10 instruments.

% \subparagraph{Environmental Sounds:} For environmental sounds, we used the FSD50K dataset \cite{fonseca2022fsd50kopendatasethumanlabeled} which contains 51197 samples for 200 different environmental sound classes. 

% \subparagraph{Speech:}
% For speech, we used the TensorFlow Speech Commands dataset \cite{speechcommands}, designed to train and evaluate keyword spotting systems. It includes 35 target words and multiple audio samples for each target class. 

\section{Datasets}

To comprehensively evaluate an LLM's audio synthesis competence, we used three datasets:
\begin{itemize}[itemsep=1pt, topsep=4pt,leftmargin=10pt]
    \item \textbf{Musical Notes:} The NSynth dataset \cite{nsynth} contains 305,979 musical notes across 10 instruments, 3 sources, and varying pitch and velocity. We sampled 1,500 audio files from 10 instruments for evaluation.
    \item \textbf{Environmental Sounds:} The FSD50K dataset \cite{fonseca2022fsd50kopendatasethumanlabeled} includes 51,197 samples from 200 environmental sound classes.
    \item \textbf{Speech:} The TensorFlow Speech Commands dataset \cite{speechcommands} provides audio samples for 35 target words, designed to train and evaluate keyword spotting systems.
\end{itemize}

\section{Methodology and Results}

In this section, we describe the experimental setup, the prompting process and prompt refinement, and the evaluation methods used.
Each experiment is designed to probe specific aspects of auditory understanding and synthesis within the selected models. By systematically evaluating these capabilities, this work aims to uncover the strengths and limitations of text-based LLMs in handling auditory information.

\subsubsection*{Representing audio with code}
For our research utilizing large language models (LLMs) not specifically trained for audio tasks, we represent audio through code. 

Although LLMs can output byte values to create audio, their capabilities are limited. The code provides a descriptive representation of the audio and captures higher-level attributes. In addition, LLMs are trained on various code examples. These models were tested using Python as the programming language.

\subsubsection*{Choice of Models}
For this study, we evaluated a range of state-of-the-art language models, including OpenAI's GPT-4 as well as Meta's Llama 3.1 70B.

\subsection{Musical Notes}
To facilitate musical notes generation, we sampled 1,500 audios from the NSynth Musical Notes dataset, ensuring both computational feasibility and diversity in the audio data. The selection process aimed to include a balanced representation of all classes, where each class corresponded to a unique combination of an instrument (bass, brass, flute, guitar, keyboard, mallet, organ, reed, string, or vocal) and a source type (acoustic, electronic, or synthetic). This resulted in 16 distinct classes, of which 12 classes contained 110 examples each. Exceptions included electronic guitar with 92 examples and acoustic flute, acoustic vocal, and electronic bass, each with fewer than 60 examples, attributable to a lower quantity of data for these classes in the NSynth dataset.

In addition to the instrument and source, each sampled audio file was annotated with its MIDI pitch, sound velocity, and a qualitative description of the note. These descriptions included characteristics such as \textit{"waveshaping that produces a distinctive crunchy sound with the presence of many harmonics"} for distorted audio or \textit{"a loud non-harmonic sound at note onset}" for percussive audio. This metadata was provided as additional information on the audio to be generated by the LLM as a prompt to aid in the synthesis of musical notes.
\subsubsection{Experiments}
Our experimentation pipeline is described in Figure \ref{fig:part_1}, where we iteratively prompt the model with detailed metadata of each of the 1500 musical notes to generate Python code that, when compiled, would reproduce the musical note as a .WAV file. We used descriptions to encapsulate the nuances of musical notes in a textual format to aid the text-based language model in generating code with the quantitative and qualitative characteristics of the desired musical note.

The experiments resulted in the successful generation of 1163 audio files by Llama 3.1-70b and 919 audio files by GPT-4. Instances of code generation failure were observed, primarily attributable to the inclusion of Python modules that were either deprecated or nonexistent within the execution environment. Analysis of error logs revealed recurring patterns, such as the generation of incorrect module names, including \textit{midutil} instead of the correct \textit{MIDIutil} and \textit{pydsmid} instead of \textit{pydsmidi}.
\subsubsection{Evaluation}
To evaluate the resulting audio, we used the Fréchet Audio Distance (FAD) \cite{kilgour2019frechetaudiodistancemetric}, a distance-based metric that measures the distributional similarity of feature representations extracted from generated and ground-truth audio. 
%For analyzing the semantic alignment between audio and audio descriptions, we initially considered CLAP (Contrastive Language-Audio Pretraining), a model designed to learn joint embeddings of audio and text. 
%Given a CLAP model’s lack of training on musical notes, we found CLAP (see section \ref{sec:evaluating_env_sounds}) to be unsuitable for retrieving accurate text embeddings for the audio descriptions. 
%Unlike environmental sounds which are inherently subjective and variable, musical notes are objective and reproducible
%: Two audios of an acoustic keyboard playing the note B2 are going to have a high degree of acoustic similarity in their spectral content, regardless of their source of production, but two dog bark sounds will be acoustically dissimilar when originating from different sources. Therefore, the use of an evaluation metric like FAD which compares ground truth audio to generated audio was better suited for the evaluation of musical notes. \\
We used CNN-based VGGish \cite{vggish2017} to extract features from each audio file. FAD between two VGGish audio features is then calculated.
Additionally, FAD was chosen as it has a high correlation with human perception of audio and because it captures the overall distributional characteristics of the generated audio embedding rather than point-wise differences. \\
The resulting FAD scores were distributed across a wide range from as low as 1.44 (mallet) to as high as 200 (bass). With no prior benchmarks for FAD scores, we employed human evaluation to evaluate a sample of generated and corresponding ground truth audios for each class at different FAD scores. The human evaluators were provided with generated audio from all the instruments across a range of FAD scores: 5, 10, 15, 20, and 25 and asked to categorize into the following categories:
\begin{itemize}[itemsep=1pt, topsep=4pt]
    \item \textbf{Highly similar}: Matches most characteristics of ground truth audios  like pitch and instrument
    \item \textbf{Moderately similar}: Matches certain sound characteristics, such as pitch, but lacks the instrument’s unique sound
    \item \textbf{Significantly distinct}: Fails to capture most characteristics of generated ground truth audio
\end{itemize}
On aggregating the results of the human evaluation, generated musical notes with an FAD score below 10 were categorized as \textbf{highly similar}, generated audios with FAD scores greater than 10 but less than 15 were categorized as \textbf{moderately similar}, and musical notes with FAD scores greater than 15 were \textbf{significantly distinct} from the ground truth.\\
The median FAD scores for 6 out of 10 instruments lie in the moderately similar category. In this category, the generated audios replicate the quantitative properties of sound accurately, like frequency, amplitude, etc. while it is unable to replicate the qualitative characteristics of sound, such as dynamic variations, to a high degree of accuracy.

\begin{figure}
    \centering
    \includegraphics[width=1\linewidth]{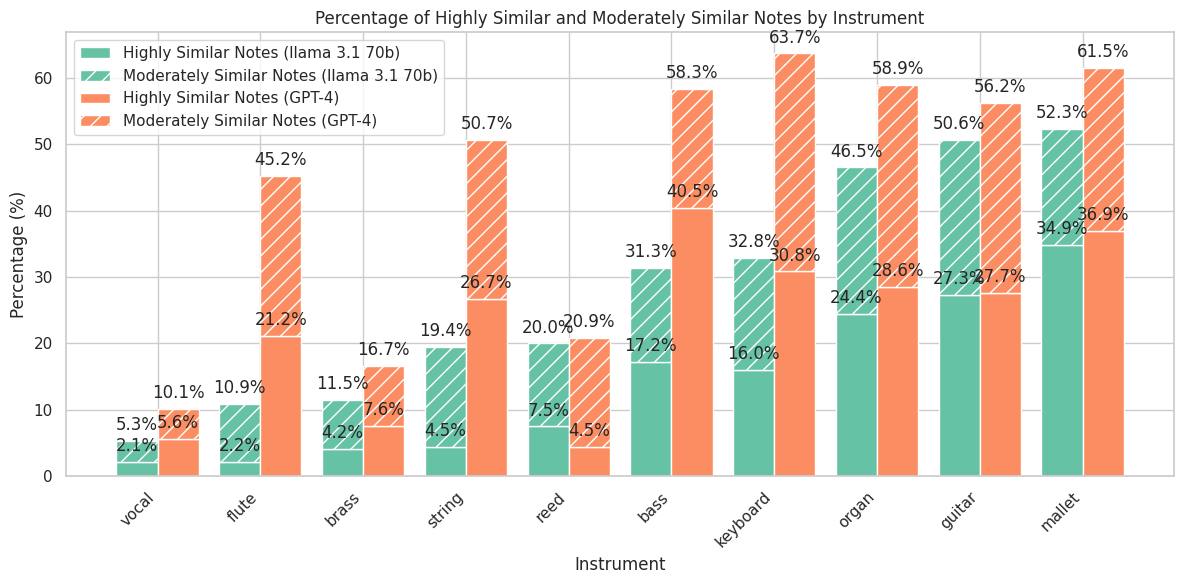}
    \caption{FAD evaluation results for Musical Notes}
    \label{fig:avg_music_eval}
\end{figure}

In our evaluation of LLM-generated musical notes, as represented in Figure \ref{fig:avg_music_eval}, musical notes generated using GPT-4 almost always outperformed notes generated by Llama 3.1-70b. 

%Percussion and keyboard instruments (mallet, keyboard, organ) consistently demonstrated the highest performance, with 61.5\%, 63.7\%, and 58.9\% of the notes generated by GPT-4 classified as at least moderately similar, respectively. Similarly, Llama 3.1-70b achieved 52.3\%, 32.8\%, and 46.5\% for the same instruments. This indicates a strong alignment between the generated audio and the target instrument characteristics in this category.
%Stringed instruments (string, bass, guitar) exhibited moderate performance, achieving 50.7\%, 58.3\%, and 56.2\% of moderately or highly similar notes with GPT-4 and 19.4\%, 31.3\%, and 50.6\% with Llama 3.1-70b. 
As seen in Figure \ref{fig:avg_music_eval}, percussion and keyboard instruments (mallet, keyboard, organ) consistently demonstrated the highest performance, followed by string instruments (string, brass, guitar). 
In contrast, wind instruments (flute, brass, reed, vocal) with the exception of flute generally performed comparatively poorly. 

\subsubsection{Comparing with GPT-4o}
\begin{figure}[t]
    \centering
    \includegraphics[width=1\linewidth]{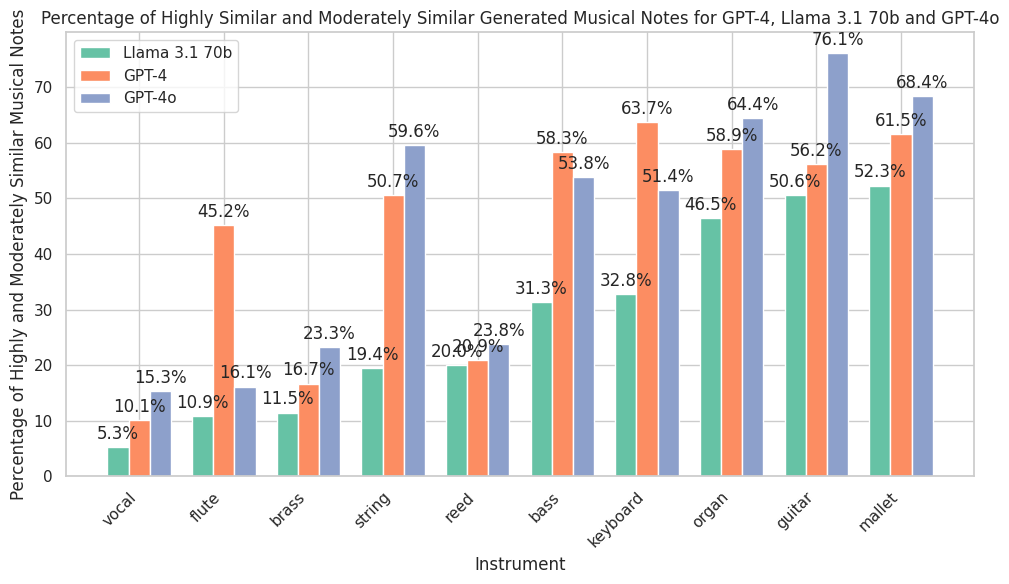}
    \caption{Percentage of Highly Similar Generated Musical Notes for GPT-4, Llama 3.1 70b and GPT-4o
}
    \label{fig:musical_notes_percentage_chart}
\end{figure}

To assess the performance of text-only models, we compared our results to OpenAI's GPT-4o, a multimodal language model trained on audio. GPT-4o outperformed the text-only models in generating musical notes for 7 out of 10 instruments and even then, only by a small amount as shown in Figure \ref{fig:musical_notes_percentage_chart}. This small difference underscores the ability of text-only models to generate musical notes almost as accurately as models trained explicitly on audio. 
%These results suggest that string instruments, while not performing as strongly as percussion and keyboard instruments, still demonstrate a reasonable capacity for similarity in generated notes, particularly for bass and guitar.

%In contrast, wind instruments (flute, brass, reed, vocal) generally performed comparatively poorly, with the combined percentage of moderately and highly similar audios falling below 25\% for both models. 
%An exception to this trend is the flute, where 45.2\% of the notes generated by GPT-4 were classified as either moderately or highly similar.

% \begin{figure}
%     \centering
%     \includegraphics[width=1\linewidth]{fad_score_per_instrument_scaled_threshold.png}
%     \caption{FAD score distribution per instrument for GPT-4}
%     \label{fig:avg_music_eval}
% \end{figure} 

\subsection{Environmental Sounds}

As the next step, we increase the complexity of the problem by prompting the model to generate environmental sounds. Examples include: an alarm clock, a bell, knocking on a door, etc. For environmental sounds, we use the Freesound Dataset 50K (FSD50K) dataset which consists of 51,197 Freesound clips unequally distributed in 200 classes drawn from the AudioSet Ontology. It consists mainly of sound events produced by physical sound sources and production mechanisms, including human sounds, sounds of things, animals, natural sounds, musical instruments, and more. The unequal distribution of audio files in each class explains that environmental sounds can have multiple versions but still belong to the same category. Dogs bark differently based on breed and other factors, but humans can distinguish between them. This introduces subjectivity and variability to the task, making it harder to generate audio compared to musical notes. We used 200 target classes and prompted the model to generate audio .WAV files for them.

\subsubsection{Experiments}
We follow the approach described in Figure \ref{fig:part_1}, similar to musical notes generation wherein we asked the model to generate code that, when compiled, generates audio representing the target class. While NSynth Musical Notes provided metadata about the audio files (used to prompt the models during generation), Freesound Dataset does not provide any useful metadata about the audio files for generation. We implement two different prompting methods to address this issue. For the first approach, we provided a \textbf{Detailed Prompt} explaining key characteristics of the output code such as (structure of the code), what parameters to consider (audio frequency, audio duration), etc. An example of the prompt is provided in the Appendix \ref{sec:appendix}.

% \begin{figure}
%     \centering
%     \includegraphics[width=1\linewidth]{download (1).png}
%     \caption{Generated spectrogram of motorcycle sound}
%     \label{fig:vroom_sound}
% \end{figure}

For the second prompting method, which we call the \textbf{Detailed Prompt + Description}, we passed a list of 200 target classes to the GPT-4o model to describe each class in detail based on its class name. We append this description to the input placeholders in the detailed prompt to provide class-specific qualitative descriptions. An example of the description of a class is shown in the Appendix \ref{sec:appendix}.

We tested these 2 generation methods on GPT-4 and Llama 3.1-70b. Each model was tasked with generating 400 samples, 200 for each method. GPT-4 generated 176/200 .WAV files (achieved an 88\% code compilation success rate) using the Detailed Prompt method and 164 .WAV files (82\%) using the Detailed Prompt + Description method whereas Llama 3.1-70b generated 98 .WAV files (49\%) using the Detailed Prompt method and 95 .WAV files (47.5\%) using the Detailed Prompt + Description method. 
\ref{tab:generated_samples_with_borders}
Similarly to the generation of musical notes, instances of failure during code execution were observed. An investigation revealed that these failures were caused by execution errors in the code due to incorrect and deprecated Python modules.

\subsubsection{\textbf{Evaluation}}
\label{sec:evaluating_env_sounds}
We evaluate the generations with the CLAP (Contrastive Language-Audio Pre-training) model. We diverge from FAD scores used in musical note generation because environmental sounds are dynamic, each class can have multiple audio files that are all correct. Additionally, the goal is to generate audio from text and evaluate how closely the audio matches the text. A shared embedding approach is strong for this task.

On comparing \textbf{Detailed Prompt} method with \textbf{Detailed Prompt + Description} method we find that Llama 3.1-70b provides more accurate results while there is not a lot of impact on GPT-4. Both models were prompted with 200 target classes using both methods.

% \begin{table*}[h]
%     \centering
%     \begin{tabular}{|l|p{3cm}|p{3cm}|p{3cm}|p{3cm}|}
%         \hline
%         \textbf{Model} & \textbf{Prompt technique} & \textbf{Total number of samples generated} & \textbf{Total number of samples correctly classified} & \textbf{Mean confidence on correctly classified generation} \\ \hline
%         \multirow{2}{*}{GPT-4} & Detailed prompt & 176 & 42 & 0.82 \\ \cline{2-5}
%         & Detailed prompt + description & 164 & 36 & 0.83 \\ \hline
%         \multirow{2}{*}{Llama 3.1-70b} & Detailed Prompt & 98 & 16 & 0.72 \\ \cline{2-5}
%         & Detailed prompt + description & 95 & 25 & 0.77 \\ \hline
%     \end{tabular}
%     \caption{Summary of results from different models and prompts techniques.}
%     \label{tab:generated_samples_with_borders}
% \end{table*}

A total of 533 audio samples were generated and the CLAP model was used to assess the similarity between the generated audio file and the target classes. The experiment was designed such that the CLAP model evaluated the similarity of the embeddings for the generated audio file against five classes. Four of which were randomly selected from the 199 classes, and the fifth was the target class. Table \ref{tab:generated_samples_with_borders} (see the appendix section \ref{sec:environmental_sounds_appendix}) summarizes the total number of generations by each model using each method, the number of generations that CLAP identified, and the mean confidence of them. The evaluation results show that GPT-4 successfully generated code that, when compiled, produced an audio file. In contrast, the code generated by Llama 3.1-70b contained numerous bugs and non-existent imports, making it challenging to compile (failed 90\% of the time). Although both models generated audio that appeared to capture some fundamental attributes of the respective class, GPT-4 demonstrated superior control over background noise, while Llama 3.1-70b generated audio with significant background noise.

Of 533 generated samples, CLAP found 119 (22.5\%) of the generated audio that belongs to the target class. A deep dive into this is shared in the Appendix \ref{sec:appendix}.

Close observation revealed that even the samples that were incorrectly classified had some distinguishable characteristics that represent the audio class that could not be identified by CLAP.

\subsubsection{\textbf{Comparing with GPT-4o}}
Our investigation into environmental sound generation paralleled our earlier work with musical notes. When prompting GPT-4o to generate environmental sounds, the results demonstrated performance metrics comparable to those of GPT-4, with no significant improvements observed. These findings suggest that LLMs may be approaching their limitations with increased complexity inherent in environmental sound generation.
%To assess the performance of text-only models, we compared our results to OpenAI's GPT-4o, a multimodal language model trained on audio. GPT-4o outperformed the text-only models in generating musical notes and environmental sounds, particularly when using detailed prompts and descriptions.

%However, our findings suggest that advancements in prompting and fine-tuning can significantly improve the performance of text-only models in generating musical notes. Additionally, the ability of text-only models to generate simple audio content raises questions about the cost-effectiveness and generalization capabilities of multimodal model training.

%While text-only models struggle with the complexity of environmental sound generation, their potential to generate basic audio content is a promising step towards more advanced cross-modal capabilities.

\subsection{Speech}
\subsubsection{Experiments}
As the final step, the complexity is elevated to its highest level by prompting the model to generate speech. We used a collection of 35 spoken target words from the Tensorflow Speech Commands dataset. Each word is represented by 1-second audio clips of a human voice uttering that particular word. We used knowledge from previous prompting methods and adapted it to fit the needs for speech generation, an example prompt can be found in the Appendix \ref{sec:appendix}. We used GPT-4, GPT-4o and Llama 3.1-70b for generation; however, due to the complexity of the task, no successful generations were observed.

% \begin{figure}
%     \centering
%     \includegraphics[width=1\linewidth]{speechgraph.png}
%     \caption{Average FAD scores for speech audio across GPT-4, O1 and Llama 3.1 70B (higher is better)}
%     \label{fig:avg_speech_eval}
% \end{figure}

% \subsubsection{Evaluation}
% \begin{itemize}
%     \item \textbf{Evaluation Metric:} We conduct a human evaluation using Mean Opinion Score (MOS) to assess the perceptual quality of the generated speech.
%     \item \textbf{Human Evaluation Setup:} A group of human evaluators is presented with the generated speech samples, as well as reference speech from the dataset. They are asked to rate the quality of each sample on a 5-point scale:
%         \begin{itemize}
%             \item 5: Excellent (indistinguishable from natural human speech)
%             \item 4: Good (natural-sounding with minimal artifacts)
%             \item 3: Fair (some noticeable artifacts but still understandable)
%             \item 2: Poor (significant artifacts, difficult to understand)
%             \item 1: Bad (unintelligible)
%         \end{itemize}
%     \item \textbf{Comparison Setup:} The evaluators compared the generated speech against a corresponding reference sample for each target word. The MOS score provides insight into the perceived quality of the generated speech regarding naturalness and intelligibility.
% \end{itemize}

\section{Conclusion}
Remarkably, text-based LLMs can understand and generate basic musical notes and environmental sounds, despite lacking direct audio training. This suggests their potential for cross-modal understanding.
However, LLMs struggle with complex instruments like wind instruments and environmental sounds due to their inherent variability and lack of detailed audio features in training data.
While LLMs cannot perfectly replicate environmental sounds, they demonstrate a basic understanding of their acoustic properties. This raises intriguing questions about how LLMs interpret textual data and the extent to which text can represent other modalities.
This research explores the extent to which text-based LLMs can understand and generate environmental sounds. While LLMs cannot perfectly replicate these sounds, they demonstrate a basic understanding of their underlying acoustic properties.
Even without explicit audio training, LLMs can map textual descriptions to auditory attributes. This raises intriguing questions about the latent information within text and the potential for LLMs to represent and generate other modalities.

\begin{figure}
    \centering
    \includegraphics[width=1\linewidth]{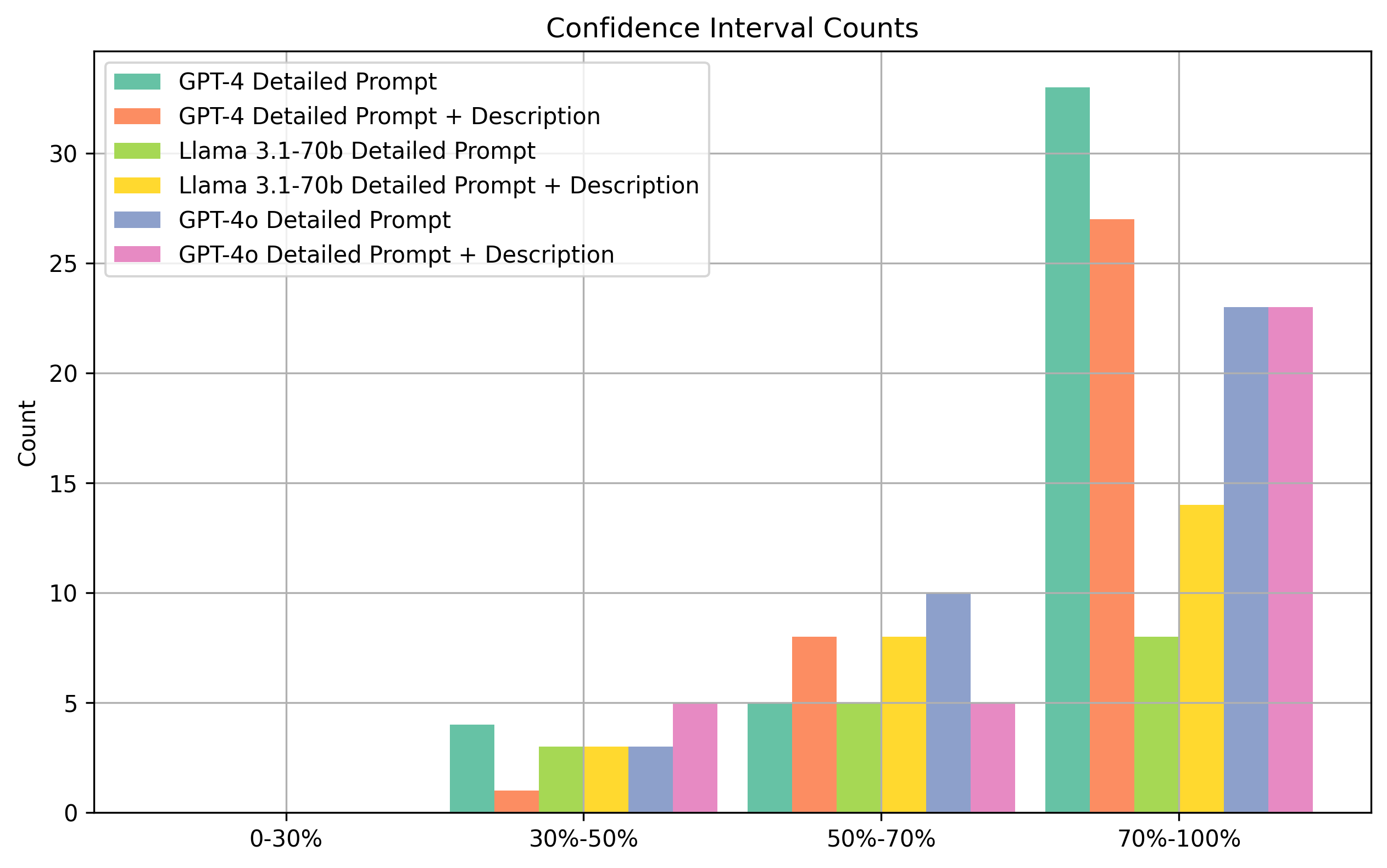}
    \caption{CLAP confidence interval in correctly predicted classes}
    \label{fig:prompt_correction_env}
\end{figure}
\bibliography{latex/sources}

\begin{thebibliography}{14}
\providecommand{\natexlab}[1]{#1}

\bibitem[{Abdou et~al.(2021)Abdou, Kulmizev, Hershcovich, Frank, Pavlick, and Søgaard}]{abdou2021languagemodelsencodeperceptual}
Mostafa Abdou, Artur Kulmizev, Daniel Hershcovich, Stella Frank, Ellie Pavlick, and Anders Søgaard. 2021.
\newblock \href {https://arxiv.org/abs/2109.06129} {Can language models encode perceptual structure without grounding? a case study in color}.
\newblock \emph{Preprint}, arXiv:2109.06129.

\bibitem[{Fonseca et~al.(2022)Fonseca, Favory, Pons, Font, and Serra}]{fonseca2022fsd50kopendatasethumanlabeled}
Eduardo Fonseca, Xavier Favory, Jordi Pons, Frederic Font, and Xavier Serra. 2022.
\newblock \href {https://arxiv.org/abs/2010.00475} {Fsd50k: An open dataset of human-labeled sound events}.
\newblock \emph{Preprint}, arXiv:2010.00475.

\bibitem[{Gong et~al.(2021)Gong, Chung, and Glass}]{gong21b_interspeech}
Yuan Gong, Yu-An Chung, and James Glass. 2021.
\newblock \href {https://doi.org/10.21437/Interspeech.2021-698} {{AST: Audio Spectrogram Transformer}}.
\newblock In \emph{Proc. Interspeech 2021}, pages 571--575.

\bibitem[{Honarvar et~al.(2023)Honarvar, van~der Wilk, and Donaldson}]{honarvar2023turbulence}
Shahin Honarvar, Mark van~der Wilk, and Alastair Donaldson. 2023.
\newblock \href {https://arxiv.org/abs/2312.14856} {Turbulence: Systematically and automatically testing instruction-tuned large language models for code}.
\newblock \emph{arXiv preprint arXiv:2312.14856}.

\bibitem[{Huang et~al.(2018)Huang, Vaswani, Uszkoreit, Shazeer, Simon, Hawthorne, Dai, Hoffman, Dinculescu, and Eck}]{huang2018musictransformer}
Cheng-Zhi~Anna Huang, Ashish Vaswani, Jakob Uszkoreit, Noam Shazeer, Ian Simon, Curtis Hawthorne, Andrew~M. Dai, Matthew~D. Hoffman, Monica Dinculescu, and Douglas Eck. 2018.
\newblock \href {https://arxiv.org/abs/1809.04281} {Music transformer}.
\newblock \emph{Preprint}, arXiv:1809.04281.

\bibitem[{Kilgour et~al.(2019)Kilgour, Zuluaga, Roblek, and Sharifi}]{kilgour2019frechetaudiodistancemetric}
Kevin Kilgour, Mauricio Zuluaga, Dominik Roblek, and Matthew Sharifi. 2019.
\newblock \href {https://arxiv.org/abs/1812.08466} {Fr\'echet audio distance: A metric for evaluating music enhancement algorithms}.
\newblock \emph{Preprint}, arXiv:1812.08466.

\bibitem[{Nagrani et~al.(2017)Nagrani, Franke, and Zisserman}]{vggish2017}
Arsha Nagrani, Jörg~K.H. Franke, and Andrew Zisserman. 2017.
\newblock \href {https://research.google.com/audioset/} {Vggish: A deep learning approach to audio classification}.
\newblock In \emph{Proceedings of the IEEE International Conference on Acoustics, Speech and Signal Processing (ICASSP)}.
\newblock Accessed: 2024-12-01.

\bibitem[{Patel and Pavlick(2022)}]{patel2022mapping}
Roma Patel and Ellie Pavlick. 2022.
\newblock \href {https://openreview.net/forum?id=gJcEM8sxHK} {Mapping language models to grounded conceptual spaces}.
\newblock In \emph{International Conference on Learning Representations}.

\bibitem[{Rubenstein et~al.(2023)Rubenstein, Asawaroengchai, Nguyen, Bapna, Borsos, de~Chaumont~Quitry, Chen, Badawy, Han, Kharitonov, Muckenhirn, Padfield, Qin, Rozenberg, Sainath, Schalkwyk, Sharifi, Ramanovich, Tagliasacchi, Tudor, Velimirović, Vincent, Yu, Wang, Zayats, Zeghidour, Zhang, Zhang, Zilka, and Frank}]{rubenstein2023audiopalmlargelanguagemodel}
Paul~K. Rubenstein, Chulayuth Asawaroengchai, Duc~Dung Nguyen, Ankur Bapna, Zalán Borsos, Félix de~Chaumont~Quitry, Peter Chen, Dalia~El Badawy, Wei Han, Eugene Kharitonov, Hannah Muckenhirn, Dirk Padfield, James Qin, Danny Rozenberg, Tara Sainath, Johan Schalkwyk, Matt Sharifi, Michelle~Tadmor Ramanovich, Marco Tagliasacchi, Alexandru Tudor, Mihajlo Velimirović, Damien Vincent, Jiahui Yu, Yongqiang Wang, Vicky Zayats, Neil Zeghidour, Yu~Zhang, Zhishuai Zhang, Lukas Zilka, and Christian Frank. 2023.
\newblock \href {https://arxiv.org/abs/2306.12925} {Audiopalm: A large language model that can speak and listen}.
\newblock \emph{Preprint}, arXiv:2306.12925.

\bibitem[{Sharma et~al.(2024)Sharma, Shaham, Baradad, Fu, Rodriguez-Munoz, Duggal, Isola, and Torralba}]{sharma2024visioncheckuplanguagemodels}
Pratyusha Sharma, Tamar~Rott Shaham, Manel Baradad, Stephanie Fu, Adrian Rodriguez-Munoz, Shivam Duggal, Phillip Isola, and Antonio Torralba. 2024.
\newblock \href {https://arxiv.org/abs/2401.01862} {A vision check-up for language models}.
\newblock \emph{Preprint}, arXiv:2401.01862.

\bibitem[{Tensorflow(2017)}]{nsynth}
Tensorflow. 2017.
\newblock Nsynth dataset.
\newblock \url{https://magenta.tensorflow.org/datasets/nsynth}.

\bibitem[{Tensorflow(2023)}]{speechcommands}
Tensorflow. 2023.
\newblock Speech commands dataset.
\newblock \url{https://www.tensorflow.org/datasets/catalog/speech_commands}.

\bibitem[{Verma and Berger(2021)}]{verma2021audiotransformerstransformerarchitectureslarge}
Prateek Verma and Jonathan Berger. 2021.
\newblock \href {https://arxiv.org/abs/2105.00335} {Audio transformers:transformer architectures for large scale audio understanding. adieu convolutions}.
\newblock \emph{Preprint}, arXiv:2105.00335.

\bibitem[{Zhang et~al.(2024)Zhang, Xie, Li, Liu, Wang, Jia, Huang, Song, Luo, Zheng, Xu, Liu, Zheng, and Liao}]{zhang2024unseen}
Yuanliang Zhang, Yifan Xie, Shanshan Li, Ke~Liu, Chong Wang, Zhouyang Jia, Xiangbing Huang, Jie Song, Chaopeng Luo, Zhizheng Zheng, Rulin Xu, Yitong Liu, Si~Zheng, and Xiangke Liao. 2024.
\newblock \href {https://arxiv.org/abs/2412.08109} {Unseen horizons: Unveiling the real capability of llm code generation beyond the familiar}.
\newblock \emph{arXiv preprint arXiv:2412.08109}.

\end{thebibliography}

\section{Appendix}
\label{sec:appendix}
\noindent \textbf{Musical Notes}
\begin{verbatim}
Provide Python code that generates a 
musical note of the following
characteristics:
    pitch - {pitch}
    velocity - {velocity}
    note - {note}
    amplitude - {amplitude} The volume of
    sound on a scale of 0 to 1
    instrument - {instrument}
    method of production - {production}
    quality description - {quality_des}
\end{verbatim}

\begin{verbatim}
Save the audio as a WAV file of 4 seconds,
named \{sound\_id\}.WAV. The audio should
have a sample rate of 16kHz. The code
should produce the most realistic and
accurate musical output possible, capturing
all the nuances of the instrument, the
production method, and the quality.
Before writing the code, think carefully
about all the musical features, audio
details, pitch, velocity, characteristics
of the instrument and the note being
simulated, quality, and method of
production. Make sure the code includes a
realistic synthesis of the musical note,
including appropriate waveforms and
harmonics for the chosen instrument,
dynamic variations in volume (e.g.,
crescendos, decrescendos) to
reflect the amplitude of sound, well-
organized structure with comments
explaining each part of the code.
Refine the outputs to achieve the most
accurate and realistic result possible.
Do not include any reasoning or
explanations in your output.
\end{verbatim}

\noindent \textbf{Environmental sounds}
\label{sec:environmental_sounds_appendix}

\begin{verbatim}
Provide Python code to generate a
realistic spectrogram and .WAV file for
the sound of {input}. Follow these steps:

1. Analyze the key characteristics of
{input}'s sound:
   - Frequency range and dominant
   frequencies
   - Temporal structure (e.g., attack,
   decay, sustain, release)
   - Any unique timbral qualities or
   harmonics

2. Write code that:
   - Uses appropriate libraries (e.g.,
   numpy, scipy, matplotlib)
   - Generates a synthetic audio signal
   matching the analyzed
   characteristics
   - Applies realistic envelopes and
   modulations
   - Creates a spectrogram
   visualization
   - Saves the audio as a .WAV file

3. Ensure the output:
   - Is 2-3 seconds in duration
   - Has a sampling rate of 44100 Hz
   - Contains minimal background noise
   - Produces a clear, recognizable
   spectrogram

4. Verify that the code:
   - Has no syntax errors
   - Handles potential exceptions
   gracefully
   - Uses proper indentation and
   follows PEP 8 style guidelines

Provide detailed comments explaining
your approach and any assumptions made
about {input}'s sound properties.
Include error handling and input
validation where necessary to prevent
runtime errors.
\end{verbatim}

\begin{verbatim}
- Alarm - A loud sound or signal used to
warn of danger or to wake someone, such as
a fire alarm or burglar alarm.
- Bell - A hollow metal instrument that
produces a ringing sound when struck,
often used for signaling or as part of a
musical instrument.
\end{verbatim}
\begin{table*}[ht]
    \centering
    \begin{tabular}{|l|p{3cm}|p{3cm}|p{3cm}|p{3cm}|}
        \hline
        \textbf{Model} & \textbf{Prompt method} & \textbf{Total number of samples generated} & \textbf{Total number of samples correctly classified} & \textbf{Mean confidence on correctly classified generation} \\ \hline
        \multirow{2}{*}{GPT-4} & Detailed prompt & 176 & 42 & 0.82 \\ \cline{2-5}
        & Detailed prompt + description & 164 & 36 & 0.83 \\ \hline
        \multirow{2}{*}{Llama 3.1-70b} & Detailed Prompt & 98 & 16 & 0.72 \\ \cline{2-5}
        & Detailed prompt + description & 95 & 25 & 0.77 \\ \hline
    \end{tabular}
    \caption{Summary of results from different models and prompt methods.}
    \label{tab:generated_samples_with_borders}
\end{table*}
Deep dive into details of correctly generated audio.
\begin{itemize}[itemsep=1pt, topsep=4pt]
    \item Maximum Confidence: 1.00
    \item Minimum Confidence: 0.33
    \item Mean Confidence: 0.80
    \item Median Confidence: 0.88
    \item Number of confidence values between 0 and 0.30: 0 (0.00\%)
    \item Number of confidence values between 0.30 and 0.50: 11 (9.24\%)
    \item Number of confidence values between 0.50 and 0.70: 26 (21.85\%)
    \item Number of confidence values between 0.70 and 1.00: 82 (68.91\%)
\end{itemize}

\noindent \textbf{Speech Prompt}
\begin{verbatim}
Provide Python code
that synthesizes a highly realistic coherent 
human speech-like audio signal of a
approximately  2 seconds in duration, 
and saves it as a WAV file. 

The synthesized speech should be clear, 
distinct, and unmistakably recognized 
as human speech,  mimicking the 
characteristics of natural human speech.

The code should simulate human vocalization 
by modeling the physical and acoustic 
processes of human speech production using 
advanced signal processing techniques. 
It should accurately  represent the word 
{word} based on the following detailed 
phonetic description {description}...
\end{verbatim}

\end{document}